\documentstyle[aas2pp4]{article}

\def\simless{\mathbin{\lower 3pt\hbox
     {$\rlap{\raise 5pt\hbox{$\char'074$}}\mathchar"7218$}}}   
\def\simmore{\mathbin{\lower 3pt\hbox
     {$\rlap{\raise 5pt\hbox{$\char'076$}}\mathchar"7218$}}}   

\received{}
\accepted{To Appear in Astrophysical Journal Letters}
\slugcomment{Submitted to Ap.J.Letters., October 1998}

\lefthead{M.~M\'endez et al.}
\righthead{kHz QPOs and mass accretion rate in 4U\,1608--52}

\begin{document}

\title{Dependence of the Frequency of the Kilohertz Quasi-Periodic
Oscillations on X-ray Count Rate and Colors in 4U\,1608--52}

\author{M.~M\'endez\altaffilmark{1,2},
        M.~van~der~Klis\altaffilmark{1},
        E.~C.~Ford\altaffilmark{1},
        R.~Wijnands\altaffilmark{1},
        J.~van~Paradijs\altaffilmark{1,3},
}

\altaffiltext{1}{Astronomical Institute ``Anton Pannekoek'',
       University of Amsterdam and Center for High-Energy Astrophysics,
       Kruislaan 403, NL-1098 SJ Amsterdam, the Netherlands}

\altaffiltext{2}{Facultad de Ciencias Astron\'omicas y Geof\'{\i}sicas, 
       Universidad Nacional de La Plata, Paseo del Bosque S/N, 
       1900 La Plata, Argentina}

\altaffiltext{3}{Physics Department, University of Alabama in Huntsville,
       Huntsville, AL 35899, USA}

\begin{abstract}

We present new results based on observations carried out with the {\em
Rossi X-ray Timing Explorer} during the decay of an outburst of the
low-mass X-ray binary (LMXB) and atoll source 4U\,1608--52.  Our results
appear to resolve, at least in 4U\,1608--52, one of the long-standing
issues about the phenomenology of the kilohertz quasi-periodic
oscillations (kHz QPOs), namely, the lack of a unique relation between
the frequency of the kHz QPOs and the X-ray flux.  We show that despite
its complex dependence on the X-ray flux, the frequency of the kHz QPOs
is monotonically related to the position of the source in the
color-color diagram.  Our findings strengthen the idea that, as in the
case of Z sources, in the atoll sources the X-ray flux is not a good
indicator of $\dot M$, and that the observed changes in the frequency of
the kHz QPOs in LMXBs are driven by changes in $\dot M$.  These results
raise some concern about the recently reported detection of the orbital
frequency at the innermost stable orbit in 4U\,1820--30.

\end{abstract}


\section{Introduction}

Nearly three years have elapsed since the first kilohertz quasi-periodic
oscillations (kHz QPOs) were discovered with the {\em Rossi X-ray Timing
Explorer} (RXTE) in the X-ray flux of Scorpius X--1 (van der Klis et al.
\cite{vanderklis96}) and 4U\,1728--34 (Strohmayer, Zhang, \& Swank
\cite{strohmayer96circ}).  In the meantime, kHz QPOs have been observed
in the persistent flux of 18 low mass X-ray binaries (LMXBs; see van der
Klis \cite{vanderklis98} for a review), both in the so called Z sources
and in the atoll sources (Hasinger \& van der Klis \cite{hasinger89},
hereafter HK89).  Except in Aql X--1, which showed a single kHz peak in
its power spectrum (Zhang et al.  \cite{zhangAqlx1}), in the other 17
sources two simultaneous kHz QPO peaks have been observed.  In some
sources a third, nearly-coherent, QPO peak has been detected during
type-I X-ray bursts, with a frequency which was consistent with being
equal to one or two times the frequency separation of the kHz QPOs
observed in the persistent flux (Strohmayer, Swank, \& Zhang
\cite{ssz98}).  It was suggested that a beat frequency mechanism is
responsible for this commensurability in the QPO frequencies (Strohmayer
et al.  \cite{strohmayer96circ}; Miller, Lamb \& Psaltis
\cite{miller98}), but this interpretation is not without problems (van
der Klis et al.  \cite{vanderklisetal97a}; M\'endez et al.
\cite{mendez98c}; M\'endez, van der Klis, \& van Paradijs
\cite{mendez98b}).

The dependence of the kHz QPO frequencies on X-ray luminosity, which is
usually assumed to be a measure of the mass accretion rate, is complex.
While in a given source, on a time scale of hours, there is a good
correlation between frequency and luminosity, sources that span nearly
three orders of magnitude in luminosity, such as Sco\,X--1 and 4U
0614+09, show kHz QPOs that cover the same range of frequencies (van der
Klis \cite{vanderklis97}).  It is as if the frequency of the kHz QPO
depends on the difference between instantaneous and average luminosity
in each source rather than on the luminosity itself.

A similar effect is seen between observations of the same source at
different epochs.  On time scales of hours, frequency and X-ray flux are
well correlated, but between different epochs the source covers the same
range of frequencies even if the average flux is different by 40\,\% or
more (e.g., Aql X--1; Zhang et al.  \cite{zhangAqlx1}).

In this paper we present new results that appear to resolve the latter
of these two issues, at least in 4U\,1608--52.  We show that while on
time scales longer than $\sim 1$ day the frequency of the kHz QPOs is
not well correlated to the X-ray flux, it is very well correlated to the
position of the source in the color-color diagram.  From this result we
conclude that the observed changes in the frequency of the kHz QPOs in
4U\,1608--52 are driven by changes in the mass accretion rate, and that
the lack of correlation between QPO frequency and X-ray count rate
occurs because, as in the case of the Z sources, in atoll sources there
is no one-to-one relationship between the observed X-ray flux and the
mass accretion rate.

\section{Observations and Results}

All the observations presented here were obtained using the proportional
counter array (PCA) onboard RXTE.  We include the data of the decay of
the 1996 outburst (Berger et al.  \cite{berger96}; M\'endez et al.
\cite{mendez98a}), and of the 1998 outburst (M\'endez et al.
\cite{mendez98c}; Fig.  1 there shows a light curve of the 1998
outburst.)  The observations as well as the modes used to record the
data are described in M\'endez et al.  (\cite{mendez98c}).  We also
include here a recent Public Target of Opportunity RXTE/PCA observation
of $\sim 7.3$ ks performed on August 6 1998, 03:13 UTC.  The observing
modes for this last observation were similar to those used by M\'endez
et al.  (\cite{mendez98c}) after 1998 March 27.

We calculated count rates in 5 energy bands, $2.0 - 3.5 - 6.4 - 9.7 -
16.0$ keV, and $2.0 - 16.0$ keV, taking into account the gain changes
applied to the PCA in March and April 1996.  In a few of the
observations one or two of the five detectors of the PCA were switched
off; we only used the three detectors which were always on to calculate
these count rates.  We subtracted the background contribution in each
band using the standard PCA background model version 2.0c\footnote{The
PCA Background Estimator is available at\\
http://lheawww.gsfc.nasa.gov/users/stark/pca/pcabackest.html, which
is maintained by NASA/GSFC.}, and normalized the count rates to 5
detectors.

In Figure \ref{figcolor} we show a color-color diagram of 4U\,1608--52.
This is the first time that 4U\,1608--52 is observed to move across all
the branches of the atoll, and constitutes one of the best examples of
the color-color diagram of an atoll source.  Based on this diagram we
conclude that in 1996, as the source count rate decreased, 4U\,1608--52
gradually moved from the lower part of the banana to the island state.
In 1998, at the peak of the outburst, 4U\,1608--52 was in the upper part
of the banana, and gradually moved down to the lower part of the banana
and the island state as the count rate decreased.  In general, the count
rate is observed to increase along the track from the island state to
the banana branch, but the relation between count rate and either of the
two colors is much less clean than that between colors.

The high-time resolution data confirm this preliminary state
classification.  We divided the $2 - 60$ keV data into segments of 256 s
and 512 s, and calculated power spectra up to a Nyquist frequency of
$2048$ Hz, normalized to fractional rms squared per Hertz.  The
characteristics of the $\simless 100$ Hz part of these power spectra
changed in correlation with the position of the source in the
color-color diagram.  When 4U 1608--52 was in the upper and lower parts
of the banana, the power spectra fitted a power law below $\sim 1$ Hz
(the Very Low Frequency Noise, VLFN), and an exponentially cut-off power
law above $\sim 1$ Hz (the High Frequency Noise, HFN; see HK89).  As
4U\,1608--52 moved from the upper to the lower parts of the banana, the
fractional amplitude of the VLFN ($0.001 - 1$ Hz) decreased from $\sim
6$\,\% rms to $\sim 2$\,\% rms, while the fractional amplitude of the
HFN ($1 - 100$ Hz) increased from $\sim 1$\,\% rms to $\sim 5$\,\% rms.
In the island state the VLFN disappeared completely (the 95\,\%
confidence upper limits were $\sim 1$\,\% rms), and the amplitude of the
HFN increased further to $\sim 10 - 17$\,\% rms.  (A more detailed
analysis of the low frequency part of the power spectra will be
presented elsewhere.)

The $\simmore 100$ Hz part of the power spectra also changed in
correlation with the position of the source in the color-color diagram:
we only observed kHz QPOs when 4U\,1608--52 was at certain positions of
the color-color diagram (see Fig.  \ref{figcolor}).  For those segments
where we observed QPOs, the $2 - 60$ keV fractional amplitudes of the
lower-frequency and higher-frequency, hereafter the lower and upper QPO,
varied from 5.3\,\% to 9.1\,\% rms, and from 3.3\,\% to 8.8\,\% rms,
respectively.  We did not detect kHz QPOs in the upper part of the
banana, with 95\,\% confidence upper limits of 0.8\,\% to 4.6\,\% rms
depending on the source count rate and the assumed width of the QPO, nor
in the extreme island state (upper right corner of the color-color
diagram), with 95\,\% confidence upper limits of 3.5\,\% to 10\,\% rms.
While these upper limits strongly suggest the absence in the upper
banana of kHz QPOs as strong as those observed in the lower banana and
the island states, we cannot rule out the presence of such kHz QPOs in
the extreme island state, when the count rates were lowest.

To further characterize the dependence of the kHz QPOs on other source
parameters, we selected only those data where we detected two
simultaneous kHz QPO peaks in the power spectrum (see M\'endez et al.
\cite{mendez98c}), so that the identification of the observed kHz peaks
is unambiguous.  We divided the data in segments of 64 s and produced a
power spectrum for each segment extending from 1/64 Hz to 2048 Hz.  In
Figure \ref{figrate} we show the dependence of the frequency of the
lower QPO, $\nu_{\rm low}$, as a function of count rate for
4U\,1608--52.  This figure shows several branches, which reflect the
relation between $\nu_{\rm low}$ and count rate during individual
observations that span from $\sim$ half an hour to $\sim 8$ hours.  The
only exception is the branch at the lowest count rate where two
different observations taken 8 days apart overlap.  This figure clearly
shows that while on time scales of a few hours or less $\nu_{\rm low}$
is well correlated to count rate, on time scales greater than $\sim$ 1
day the relation is complex and $\nu_{\rm low}$ is not uniquely
determined by the count rate.  We obtained the same result using the $2
- 60$ keV source flux instead of the count rate.

There is a much better correlation between $\nu_{\rm low}$ and the
position of the source on the color-color diagram.  In Figure
\ref{fig_hc} we show $\nu_{\rm low}$ as a function of hard color (see
Fig.  \ref{figcolor}) for the same intervals shown in Figure
\ref{figrate}.  The complexity seen in the frequency vs.  count rate
diagram (Fig.  \ref{figrate}) is reduced to a single track in the
frequency vs.  hard color diagram.  The frequency of the lower kHz QPO
increases as the hard color decreases, i.e., as 4U\,1608--52 moves from
the island state to the lower banana, and keeps increasing at the turn
of the lower banana, where the hard color reaches its lowest value of
$\sim 0.40$.

The shape of the track in Figure \ref{fig_hc} suggests that the hard
color may not be sensitive to changes of state when the the source moves
into the banana in the color-color diagram.  To further investigate
this, we applied the $S_{\rm Z}$ parameterization (e.g., Wijnands et al.
\cite{wijnandsa&a97}), which we call $S_{\rm a}$ for atoll sources, to
the color-color diagram.  In this phenomenological approach we
approximated the shape of the color-color diagram with a spline (Fig.
\ref{figcolor}), and we used the parameter $S_{\rm a}$ to measure
positions along this spline.  We set $S_{\rm a}$ to 1 at (2.67,0.77),
and to 2 at (2.19,0.42), as indicated in Figure \ref{figcolor}.  We
measured the position on the color-color diagram of each of the 64-s
segments.  We then grouped these segments into 36 sets, such that within
each set $\nu_{\rm low}$ did not vary by more than $\sim 25$ Hz, and the
source count rate did not vary by more than 10\,\%.  (This is to avoid
mixing segments that come from different branches in Figure
\ref{figrate}.)  For each set we calculated the average frequency of the
lower and upper kHz QPO peak, and of $S_{\rm a}$.  In Figure
\ref{fig_sa} we plot the frequencies of both kHz QPOs as a function of
$S_{\rm a}$.  The error bars are the standard deviation of each
selection.  We also include in this figure several extra sets (open
squares) for which we only detected one of the kHz QPOs, which therefore
we could not a priori identify as the upper or lower peak.  As expected,
$S_{\rm a}$ is more sensitive than the hard color to changes of state
when the the source moves from the island state into the turn of the
banana:  $S_{\rm a}$ increases from $\sim 2.04$ to $\sim 2.15$ there,
while the hard color saturates at $\sim 0.40$.  Figure \ref{fig_sa}
confirms that the frequency of the kHz QPOs is strongly correlated to
the position of the source in the color-color diagram.  It also shows
that the steep rise of kHz QPO frequency at the turn of the lower banana
is not just due to a lack of sensitivity of the hard color to state
changes in this part of the color-color diagram, but that it occurs as a
function of the position in the color-color diagram as well.

\section{Discussion}

Previous observations of kHz QPOs in atoll sources have resulted in a
confusing picture concerning the dependence of QPO frequency on mass
accretion rate.  Cases have been reported where, in a given source, QPO
frequency showed a good correlation to count rate or spectral hardness
(Ford, van der Klis, \& Kaaret \cite{fvdkk98}; Wijnands et al.
\cite{wijnands1636}; Wijnands et al.  \cite{wijnands1735}; Ford et al.
\cite{ford1735}; Zhang et al.  \cite{zhang1820}; Wijnands \& van der
Klis \cite{wijnands1731}).  In other cases, a correlation with count
rate or flux was conspicuously lacking (Ford et al.  \cite{ford97a};
M\'endez et al.  \cite{mendez98a}; Zhang et al.  \cite{zhangAqlx1}).

\begin{figure}[ht]
\plotfiddle{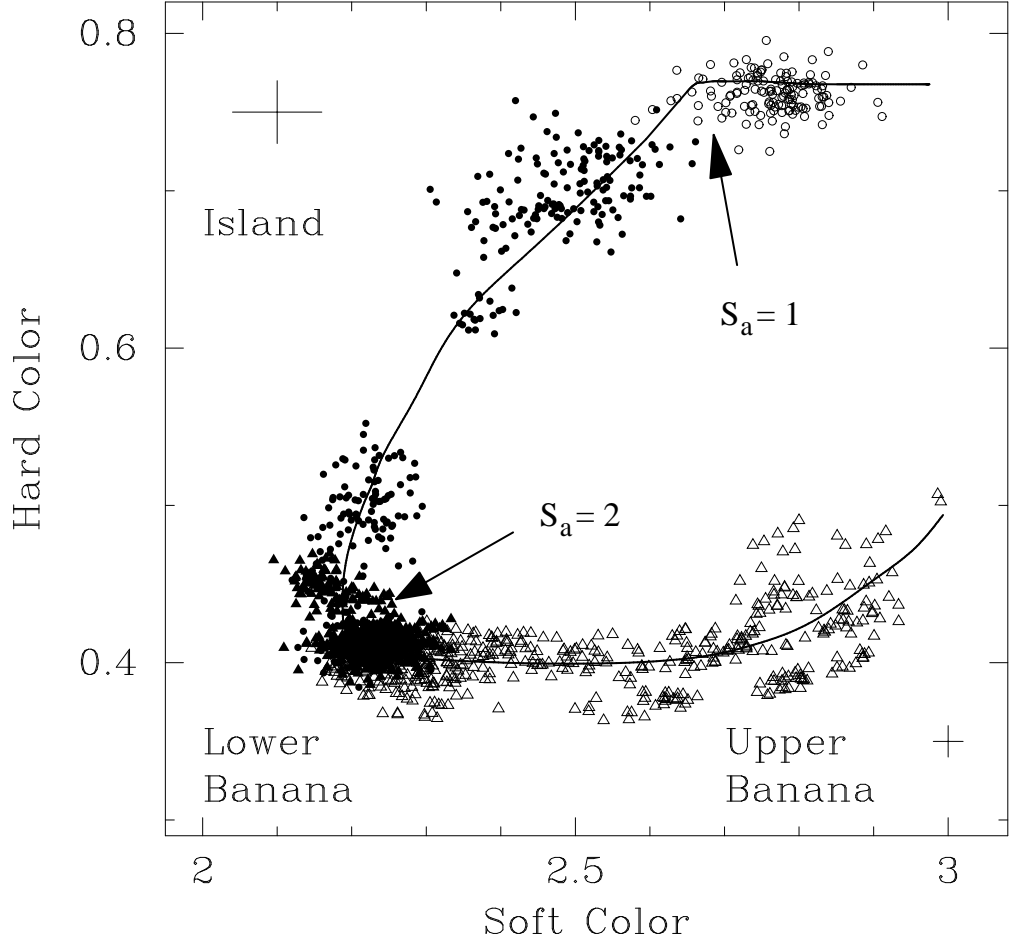}{220pt}{270}{70}{70}{-270}{250}
\vspace{6cm}
\caption{
Color-Color diagram of 4U\,1608--52.  The soft and hard colors are
defined as the ratio of count rates in the bands $3.5 - 6.4$ keV and
$2.0 - 3.5$ keV, and $9.7 - 16.0$ keV and $6.4 - 9.7$ keV, respectively.
The contribution of the background has been subtracted, but no dead-time
correction was applied to the data (in this case the dead-time effects
on the colors are less than 1\,\%). Each point represents 128 s of
data.  We show the typical error bars in the banana and the island
states.  Filled and open symbols indicate segments with and without kHz
QPOs, respectively.  We represent the island and banana states (defined on
the basis of the low-frequency part of the power spectra) by circles and
triangles, respectively.  The curve shows the parametrization of the
color-color diagram in terms of $S_{\rm a}$ (see text).
\label{figcolor}
}

\end{figure}

\begin{figure}[ht]
\plotfiddle{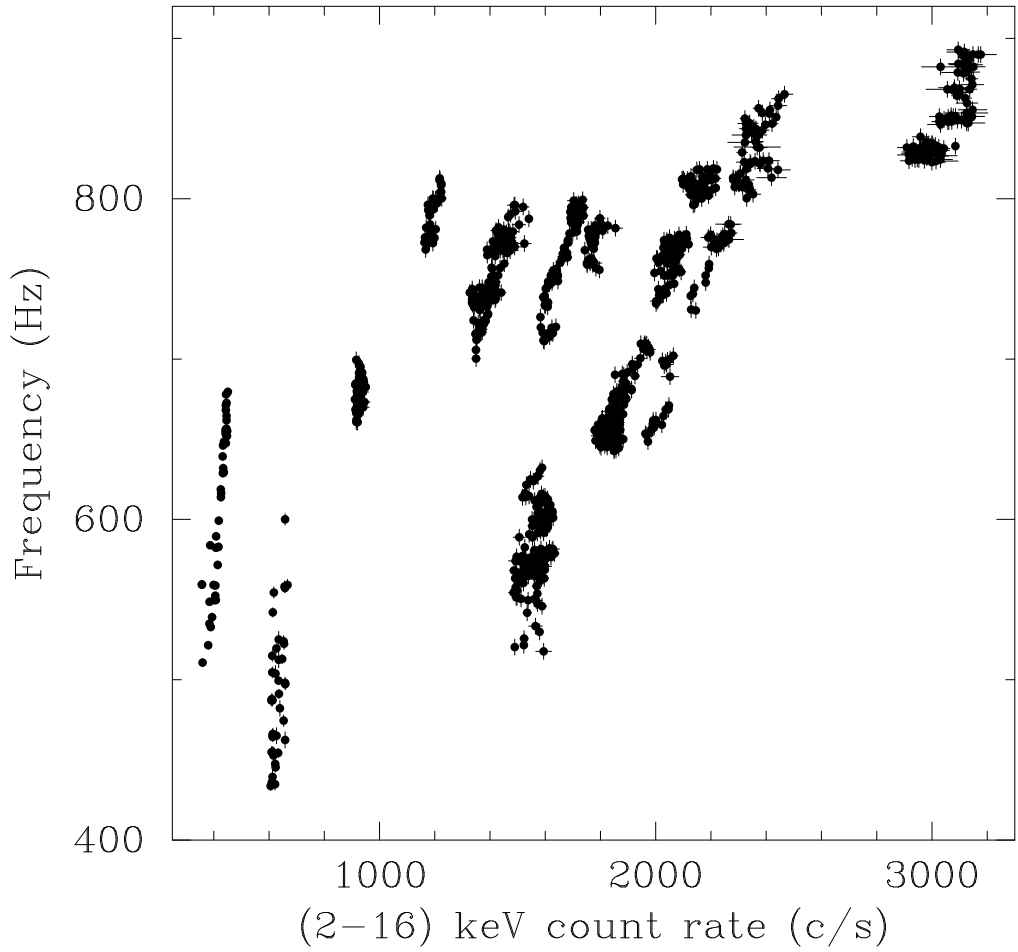}{220pt}{270}{70}{70}{-270}{250}
\vspace{6cm}
\caption{
Relation between the frequency of the lower kHz QPO and
the $2 - 16$ keV count rate. The count rates have been corrected for
background, and normalized to 5 detectors. Each point represents a
64-s segment (see text). We only include data where both kHz QPOs
are detected simultaneously , so that we can unambiguously identify
the lower kHz peak.
\label{figrate}
}

\end{figure}

\begin{figure}[ht]
\plotfiddle{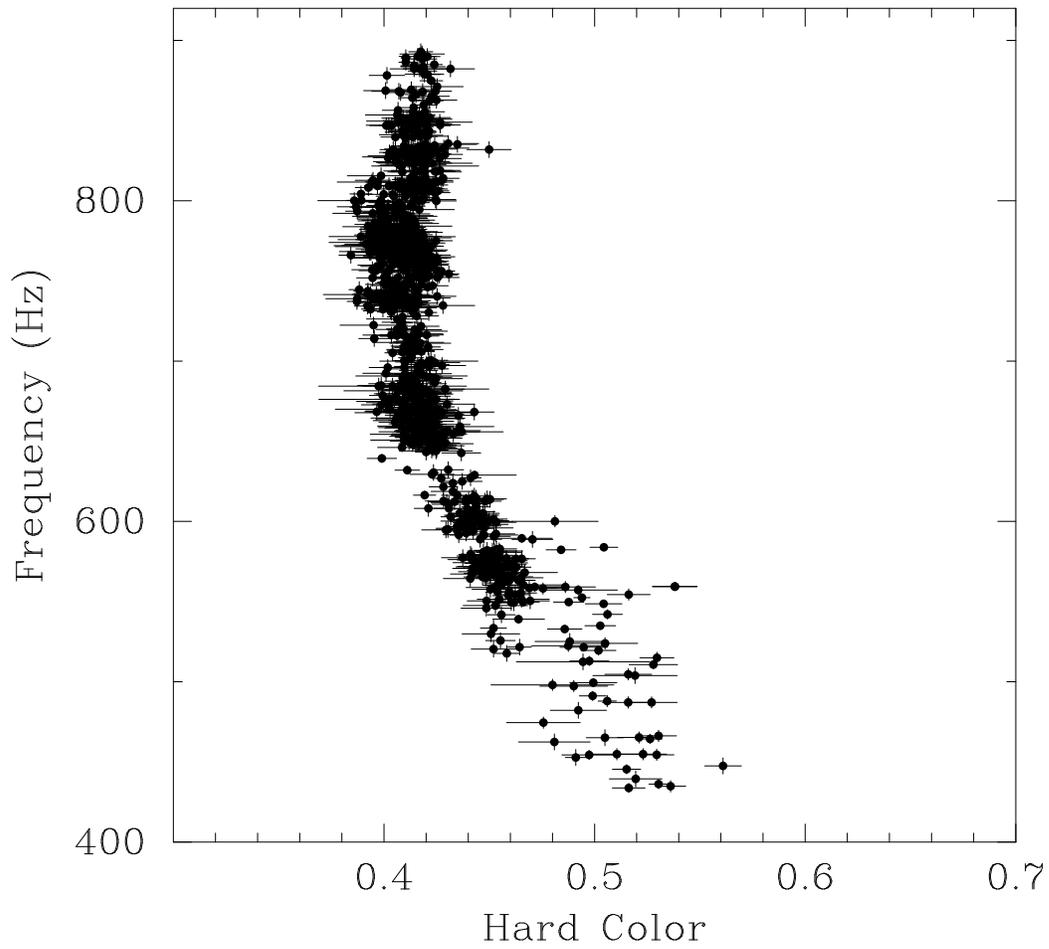}{220pt}{270}{70}{70}{-270}{250}
\vspace{6cm}
\caption{
Relation between the frequency of the lower kHz QPO and the hard color
(see Fig. 1), for the same segments shown in Figure 2.
\label{fig_hc}
}

\end{figure}

\begin{figure}[ht]
\plotfiddle{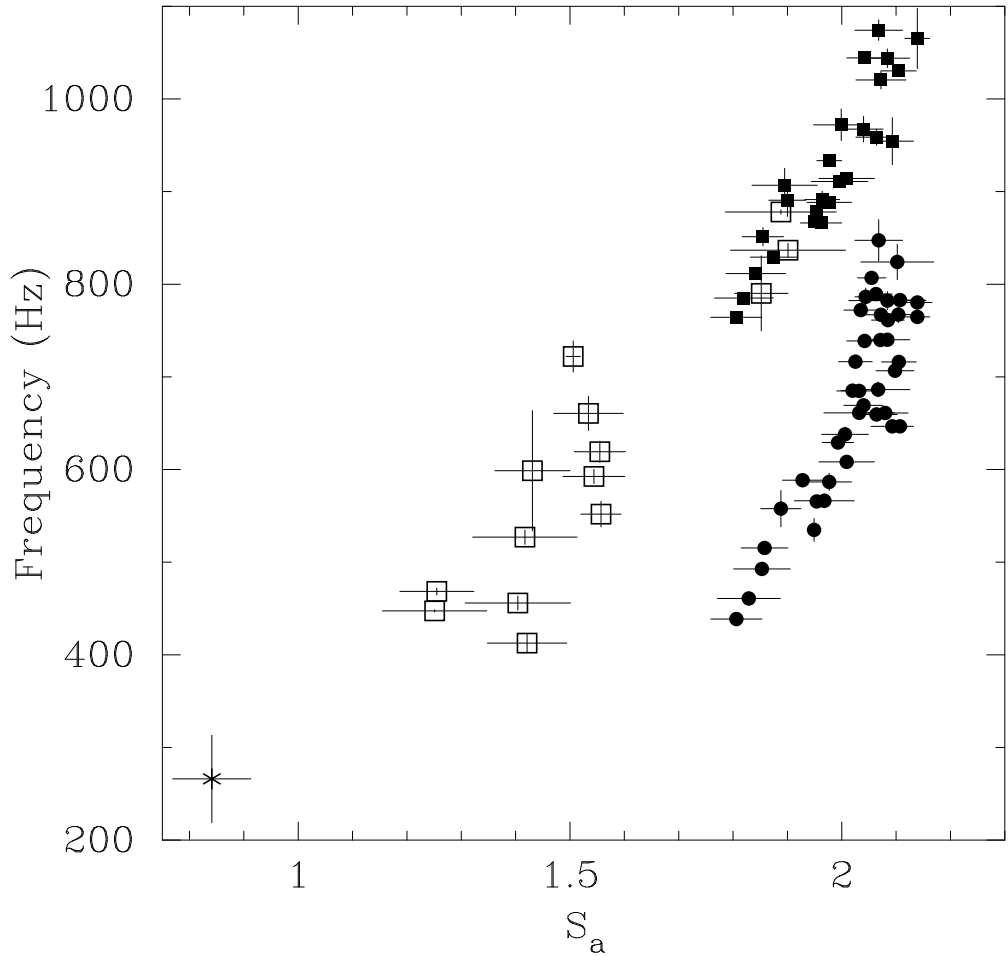}{220pt}{270}{70}{70}{-270}{250}
\vspace{6cm}
\caption{
Diagram of the frequencies of the upper and lower kHz QPOs vs.
the position of the source on the color-color diagram, as measured by
$S_{\rm a}$ (see Fig. 1, and text). Circles and filled squares
represent the lower and the upper kHz peak, respectively. Open squares
represent segments where we only detected one of the kHz QPOs, and therefore
we could not identify it as the upper or lower peak. The asterisk is
one observation where we see a single QPO at a level of significance
of $3 \sigma$ only.
\label{fig_sa}
}
\end{figure}

Our observations show for the first time that a total lack of
correlation between frequency and count rate on time scales longer than
a day (Fig.  \ref{figrate}) can coexist with a very good correlation
between frequency and position in the X-ray color-color diagram (Fig.
\ref{fig_hc}).  The frequency increases with $S_{\rm a}$, as the source
moves from the island to the banana.  Only on time scales of hours does
the QPO frequency appear to also correlate well with count rate.  The
presence of the QPOs also correlates well with the position in the
color-color diagram:  the QPOs are only detected in the lower banana and
the moderate island states, and disappear both in the upper banana and
in the extreme island states (Fig.  \ref{figcolor}).

In atoll sources $\dot M$ is thought to increase monotonically with
$S_{\rm a}$ along the track in the color-color diagram, from the island
to the upper banana (HK89), whereas X-ray count rate tracks $\dot M$
much less well (van der Klis et al.  \cite{vdk90}; van der Klis
\cite{vdk94apj}; Prins \& van der Klis \cite{prins97}).  The properties
of the $< 100 $ Hz power spectra depend monotonically on inferred $\dot
M$ (HK89).  Our result that the frequency of the kHz QPO is well
correlated to $S_{\rm a}$, but not to X-ray count rate, implies that in
4U\,1608--52 the kHz QPO frequency {\em also} depends monotonically on
inferred $\dot M$.  By extension this conclusion applies to each atoll
source; in Z sources similar conclusions were previously also reached
(e.g., Wijnands et al.  \cite{wijnands17+2}).  Our analysis, however,
sheds no light on the question why sources with very different inferred
$\dot M$ (e.g., 4U 0614+09 and Sco\,X--1) can have kHz QPO in the same
range of frequencies.

Further indications for this interpretation come from the simultaneous
analysis of the low and high frequency parts of the power spectra of
other atoll sources.  In 4U\,1728--34 the kHz QPO frequencies were
recently found to be very well correlated to several $< 100$ Hz power
spectral properties (Ford \& van der Klis \cite{ford&vdk98}), and
Psaltis, Belloni and van der Klis (\cite{psaltisetal98}) obtained a
similar result for a number of other atoll (and Z) sources.  In all
these sources, not only the position in the color-color diagram and the
various low frequency power spectral parameters, but also the
frequencies of the kHz QPOs are all well correlated with each other.
This indicates that the single parameter, inferred to be $\dot M$, which
governs all the former properties also governs the frequency of the kHz
QPO.

X-ray intensity is the exception:  it can vary more or less
independently from the other parameters.  In 4U\,1608--52, it can change
by a factor of $\sim 4$ (see Fig.  \ref{figrate}) while the other
parameters do not vary significantly.  If as inferred, this constancy of
the other parameters means that $\dot M$ does not change, then this
indicates that strongly variable beaming of the X-ray flux, or
large-scale redistribution of some of the radiation over unobserved
energy ranges is occurring in order to change the flux by the observed
factors, {\em without} any appreciable changes in the X-ray spectrum.
We may need to scrutinize more closely the concept of $\dot M$ in order
to solve this dilemma.  For example, perhaps the $\dot M$ governing all
the other parameters is the $\dot M$ through the inner accretion disk,
whereas matter also flows onto the neutron star in a more radial inflow,
or away from it in a jet.

In 4U\,1608--52 we observe no evidence for a saturation of the frequency
of the kHz QPOs at a constant maximum value as $\dot M$ increases,
different from what Zhang et al.  (\cite{zhang1820}) inferred for 4U
1820--30.  They presented data in which the kHz QPO frequencies increase
with count rate up to a threshold level, above which the frequencies
remain approximately constant while the count rate keeps increasing.
Interpreting count rate as a measure for $\dot M$ they argue that this
is evidence for the inner edge of the disk reaching the
general-relativistic innermost stable orbit.  However, we have shown
here that, at least in 4U\,1608--52, count rate is not a good measure
for $\dot M$.  Inspection of Figure \ref{figrate} suggests that with
sparser sampling our plot could easily have looked similar to the one
presented by Zhang et al.  (\cite{zhang1820}) for 4U\,1820--30.  It will
therefore be of great interest to see if in 4U\,1820--30 the saturation
of QPO frequency as a function of count rate is still there when this
parameter is plotted as a function of position in the X-ray color-color
diagram.

\acknowledgements

This work was supported in part by the Netherlands Organization for
Scientific Research (NWO) under grant PGS 78-277 and by the Netherlands
Foundation for research in astronomy (ASTRON) under grant 781-76-017.
MM is a fellow of the Consejo Nacional de Investigaciones
Cient\'{\i}ficas y T\'ecnicas de la Rep\'ublica Argentina.  JVP
acknowledges support from the National Aeronautics and Space
Administration through contract NAG5-3269, 5-4482 and 5-7382.  This
research has made use of data obtained through the High Energy
Astrophysics Science Archive Research Center Online Service, provided by
the NASA/Goddard Space Flight Center.

\onecolumn

\clearpage

\clearpage

\clearpage

\clearpage

\end{document}